\documentclass[%
 reprint,
showpacs,
 amsmath,amssymb,
 aps,
]{revtex4-1}

\usepackage{graphicx}
\usepackage{dcolumn}
\usepackage{bm}
\usepackage[left=1.5cm, right=1.5cm, top=1.785cm, bottom=2.0cm]{geometry}
\usepackage{color,xcolor,colortbl}
\usepackage{hyperref}
\usepackage[version=3]{mhchem} 
\usepackage{array}
\usepackage{zref-xr}
\usepackage{textgreek}
\usepackage{amsmath}
\usepackage{siunitx} 
\usepackage{graphicx}
\usepackage{caption}
\usepackage[labelformat=simple]{subcaption}
\usepackage{float}
\usepackage{adjustbox}
\usepackage{setspace}
\usepackage{rotating}
\usepackage{tablefootnote} 
\usepackage{booktabs,caption} 
\usepackage{booktabs}
\usepackage[flushleft]{threeparttable} 
\usepackage{multirow}
\usepackage{tabularx}
\usepackage{lastpage}
\usepackage[english]{babel}
\usepackage{epstopdf}

\begin{document}

\author{David Mora-Fonz}
\email{david.fonz.11@ucl.ac.uk}
\affiliation{Department of Physics and Astronomy, University College London, Gower Street, London WC1E 6BT, United Kingdom}
\author{Moloud Kaviani}
\affiliation{Department of Chemistry and Biochemistry, University of Bern, Freiestrasse 3, CH-3012 Bern, Switzerland}
\author{Alexander L. Shluger}
\affiliation{Department of Physics and Astronomy, University College London, Gower Street, London WC1E 6BT, United Kingdom}

\title{Disorder-induced electron and hole trapping in amorphous \ce{TiO2}}


\begin{abstract}
Thin films of amorphous (a)-\ce{TiO2} are ubiquitous as photocatalysts, protective coatings, photo-anodes and in memory application, where they are exposed to excess electrons and holes. We investigate trapping of excess electrons and holes in the bulk of pure amorphous titanium dioxide, a-\ce{TiO2}, using hybrid density functional theory (h-DFT) calculations. Fifty 270-atom a-\ce{TiO2} structures were produced using classical molecular dynamics and their geometries fully optimised using h-DFT simulations. They have the density, distribution of atomic coordination numbers and radial pair-distribution functions in agreement with experiment. The calculated average a-\ce{TiO2} band gap is \SI{3.25}{\electronvolt} with no states splitting into the band gap. Trapping of excess electrons and holes in a-\ce{TiO2} is predicted at precursor sites, such as elongated \mbox{Ti--O} bonds. 
Single electron and hole polarons have average trapping energies ($E_T$) of \SI{-0.4}{\electronvolt} and \SI{-0.8}{\electronvolt}, respectively. We also identify several types of electron and hole bipolaron states and discuss their stability. These results can be used for understanding the mechanisms of photo-catalysis and improving the performance of electronic devices employing a-\ce{TiO2} films.
  
\end{abstract}
\maketitle

\section{Introduction}

\ce{TiO2}-based materials and devices are studied extensively due to their optical, dielectric, catalytic, thermal and mechanical properties 
(see e.g. refs.~\citenum{linsebigler1995photocatalysis,pelaez2012review,schneider2014understanding,rahimi2016review} 
and references therein). In technological applications, these materials are often produced as thin films or powders by employing various techniques, such as atomic layer deposition,  chemical vapour deposition, electron beam deposition, reactive evaporation, plasma plating, sputtering and others.\cite{profijt2011plasma,niemela2017titanium} The initially grown \ce{TiO2} films and nano-particles are  amorphous or polycrystalline \cite{petkov1998atomic} and undergo further thermal treatment to achieve desired technological properties. Amorphous \ce{TiO2} films are used as protective coatings for concentrated solar power mirrors \cite{ennaceri2016deposition}, photo-anodes \cite{hu2014amorphous} and in nonvolatile memory applications.\cite{jeong2010interface} 

The performance of these as well as other systems used in photocatalysis and photoelectrocatalysis \cite{Sun2019amorphTiO2} is affected or governed by electrons and/or holes induced by dopants (such as H, Li, Nb, and vacancy defects),\cite{Valentin_copsi_2005,Valentin_rantd_2009,setvin2014direct,bogomolov1968optical,PhysRevLett.117.116402,Deskins_dotss_2011} photo-excitation and as a result of carrier injection from electrodes. In similar wide gap semiconductors excess charges often localise at regular lattice sites or impurities and modify the electronic structure by creating the corresponding shallow or deep gap states.\cite{rettie2016unravelling}  They may degrade the electronic properties of the material or provide opportunities for band gap engineering. For example, electron and hole localisation reduces electrical and photo-conductivity of materials. In contrast, electron–hole recombination is highly undesirable leading to short exciton lifetime and a poor photocatalyst. Therefore, it is important to understand how excess carriers interact with materials. 

Small polarons in the two main \ce{TiO2} polymorphs, rutile and anatase, have been studied experimentally
\cite{PhysRevB.87.125201,PhysRevLett.117.116402,eagles1964polar,macdonald2010situ,chiesa2013charge} and theoretically using DFT-based methods,\cite{austin2001polarons,sezen2014probing,deak2011polaronic,deak2012quantitative,zawadzki2011electronic,janotti2013dual,tabriz2017application,spreafico2014nature,selccuk2017excess,Elmaslmane_TiO2_polaron18,Deak_qtoto_2012,Morgan_ptoea_2009,Deskins_etvph_2007} and their properties have recently been reviewed in e.g. refs.~\citenum{lany2015semiconducting,Elmaslmane_TiO2_polaron18,Reticcioli2019polaronhandb} and references therein. 
In rutile, there is a general agreement that self-trapped electron polarons are stable: experiments report a much higher small electron polaron transport (with thermal activation energies around $-20$ and \SI{-30}{\milli\electronvolt})\cite{PhysRevB.87.125201} than a band-like conduction, which is an indication of the preference for localised electrons; whereas calculations predict exothermic $E_T$ ranging from $-0.02$ to \SI{-0.4}{\electronvolt}.\cite{lany2015semiconducting,tabriz2017application,Elmaslmane_TiO2_polaron18} For self-trapped holes in rutile, opinions are divided, with 
electron paramagnetic resonance measurements suggesting either the formation \cite{Yang_psthc_2010} or no formation of hole traps.\cite{macdonald2010situ} Similarly, DFT (DFT+$U$,  DFT + polaron correction \cite{lany2015semiconducting}) and  and h-DFT calculations (HSE06, PBE0-TC-LRC \cite{Elmaslmane_TiO2_polaron18}) predict exothermic and endothermic hole $E_T$, respectively. 
In anatase, both experiment and theory agree that holes are trapped in deep hole polaron states, but electron polarons are metastable. The experimental evidence shows that the hole trapping in anatase is deeper and more localised than the electron trapping in rutile.\cite{berger2005light}  


In spite of wide applications, relatively little is still known regarding intrinsic electron and hole trapping in a-\ce{TiO2}. The structure and electronic properties of stoichiometric and reduced a-\ce{TiO2} have been modeled using a combination of classical structure simulations and DFT-based calculations.\cite{pham2015oxygen,deskins2017structural,landmann2012fingerprints,prasai2012properties}  DFT+$U$ calculations \cite{pham2015oxygen} of a single amorphous structure predicted the hole $E_T$ at intrinsic sites in a-\ce{TiO2} to be much larger than in rutile \SI{+0.5}{\electronvolt} and of the order of \SI{-1.3}{\electronvolt}. 
Besides, trapping of electrons has been observed in the bulk of a-\ce{TiO2} nanoparticles \cite{rittmann2014mapping} and electron trapping induced by Fe impurities in a-\ce{TiO2} was calculated using DFT+$U$.\cite{ghuman2016self}

 Deep electron and hole trapping has been predicted in several other amorphous oxides where polarons do not trap or form only shallow states in crystalline phase. \cite{Strand2018JPCM} In particular, theoretical studies have shown that the precursor sites composed of wide O-Si-O bond angles act as deep electron traps in amorphous \ce{SiO2}.\cite{el2014nature} These sites can accommodate up to two extra electrons.\cite{gao2016mechanism} In amorphous \ce{InGaZnO4} \cite{nahm2014undercoordinated} and \ce{HfO2} \cite{kaviani2016deep} under-coordinated indium and hafnium atoms, respectively, are shown to serve as precursors for the deep electron trapping. Similar precursors also act as electron traps at surfaces and grain boundaries.\cite{wallace2014grain,di2011bulk,wallace2015facet,yamamoto2012hybrid,morgan2007DFT+U} Holes have been shown to trap at low-coordinated oxygen sites in most amorphous oxides. \cite{Strand2018JPCM,Mora-Fonz_moiea_2020}
 
 The fact that disorder can present precursor sites for formation of deep localised electron and hole states suggests that balance between the short-ranged phonon-mediated attraction and on-site Hubbard repulsion could be tipped in favour of formation of bipolaron like states where two electrons or holes co-exist on one or several neighbouring network sites. The possibility of formation of such states in crystals \cite{Cohen1984bipolaroncryst,Wellein1996bipolaron} and amorphous solids \cite{Andersen1975polaron,Cohen1984disorder} has been predicted by the phenomenological theory and demonstrated in some oxides by DFT simulations. \cite{Strand2018JPCM} The electron bipolarons have been predicted by DFT calculations in amorphous \ce{SiO2} \cite{gao2016mechanism} and \ce{HfO2} \cite{kaviani2016deep}. Hole bipolarons have been predicted in both crystalline (e.g. \ce{BiVO4} \cite{Ambrosio2019BiVO4}, \ce{V2O5}, \ce{TiO2} \cite{Chen2014doublehole}) and amorphous oxides (\ce{Al2O3}, \ce{HfO2} \cite{Strand2017holes} and \ce{TiO2} \cite{Guo2018holediff}). The formation of hole bipolaron states is associated with the formation of peroxide-like O--O states inside the oxide. \cite{Chen2014doublehole,Strand2017holes} 

 Here, we study the electronic structure and electron and hole trapping properties of a-\ce{TiO2} using a h-DFT approach, which has been carefully calibrated to model polarons in six different phases of \ce{TiO2} in ref.~\cite{Elmaslmane_TiO2_polaron18}. Fifty 270-atom structures were used to obtain the distribution of structural and electronic properties of a-\ce{TiO2}. 
Every structure was tested for electron and hole trapping using the inverse participation ratio (IPR) analysis. We demonstrate that 
Ti and O ions serve as precursor sites for deep electron and hole trapping in a-\ce{TiO2}  with an average $E_T$ of about \SI{-0.8}{\electronvolt} and \SI{-0.4}{\electronvolt} for holes and electrons, respectively. We also identify several types of electron and hole bipolaron-type states and discuss their stability. These results may have important implications for applications for understanding the properties and performance of (photo)catalysis, electronics, memory devices and batteries.

\section{Computational Details}

The structure of a-\ce{TiO2} has been studied experimentally in refs.\cite{petkov1998atomic,tanaka2001effects,ottermann1996young}. Following the success in experimental preparation of metastable metal alloys \cite{Duwez1960alloy}, theoretical models of oxide glasses are also usually obtained using a melt-quench procedure and molecular dynamics (MD) \cite{Vollmayr1996SiO2}. This technique has been used to model structures of amorphous a-\ce{HfO2}, a-\ce{SiO2}, a-\ce{Al2O3}, a-\ce{ZnO} and a-\ce{Sm2O3} 
 \cite{kaviani2016deep,el2014nature,dicks2017theoretical,Mora-Fonz_maztp_2019,Mora-Fonz_moiea_2020,Olsson_sevae_2019} as well as other non-glass forming oxides.\cite{Medvedeva2017rev} Similarly, classical force-fields, \cite{matsui1991molecular,van2007structural,pham2015electronic,pham2015oxygen,hoang2007structural,ghuman2013effect,van2007pressure,van2007structural,lumpkin2008experimental,kaur2011structure}
Density Functional based Tight Binding (DFTB)~\cite{kohler2013computational} 
and DFT \cite{prasai2012properties,landmann2012fingerprints,Guo2018holediff} simulations have been used to create models of a-\ce{TiO2} structures.

We employed classical force-field MD simulations followed by a complete structural relaxation of obtained structures using h-DFT. We used the LAMMPS package \cite{plimpton1995fast} with the Matsui-Akaogi force field \cite{matsui1991molecular} which has been shown to reproduce the structural properties of the crystalline, liquid and the amorphous phases of \ce{TiO2} with accuracy close to the first-principle methods.\cite{kohler2013computational,landmann2012fingerprints} To study the distribution of properties of trapped carriers in a-\ce{TiO2}, we created fifty different amorphous structures using MD at constant pressure and a Nosé-Hoover thermostat and barostat.  
As the initial structure, in all cases, we used a cubic periodic cell containing  270 atoms distributed randomly across the simulation cell. First, the structures were equilibrated at \SI{300}{\kelvin} for \SI{50}{\pico\second} and then the temperature was linearly increased to \SI{5000}{\kelvin} for \SI{50}{\pico\second}. The melt was further equilibrated for \SI{500}{\pico\second} at \SI{5000}{\kelvin}. The systems were cooled down from \SI{5000}{\kelvin} to \SI{300}{\kelvin} during \SI{4.7}{\nano\second} with a cooling rate of \SI{1}{\kelvin\per\pico\second}. Finally, the structures were equilibrated for \SI{50}{\pico\second} at \SI{300}{\kelvin}. We note that the initial structure has no effect on the topology of our amorphous structures due to the long-time simulation of the melt.

Further optimisation of the geometry and  volume of these structures along with the subsequent electronic structure calculations were performed using DFT as implemented in the CP2K code.\cite{vandevondele2005quickstep,guidon2009robust} It employs a Gaussian basis set mixed with an auxiliary plane-wave basis set.\cite{lippert1997hybrid}  The double- and triple-$\zeta$ Gaussian basis-sets \cite{vandevondele2007gaussian} were employed on oxygen and titanium atoms in conjunction with the GTH pseudopotential.\cite{goedecker1996separable} The plane-wave cutoff was set to \SI{8163}{\electronvolt} (600 Ry), which is sufficient to converge the r~-~\ce{TiO2} bulk lattice energy (6 atoms) to less than \SI{1}{\milli\electronvolt}. To avoid the bond-length overestimation typical for GGA functionals, for preliminary geometry optimization we used the PBEsol functional \cite{PhysRevLett.100.136406} which is a flavor of the well-known PBE functional~\cite{perdew1996generalized} and is known to produce lattice parameters in solids with relatively higher accuracy. This gives a better starting point for the subsequent more expensive hybrid functional calculations. 

Accurate prediction of polaron states is challenging due to the self-interaction error inherent in DFT.\cite{Cohen_cfdft_2012,Gavartin_mcsti_2003,Pacchioni_tdohl_2000,Laegsgaard_htaai_2001} It is widely accepted that hybrid functionals, can reliably describe properties of insulators and currently present the best choice to accurately describe localised electron and hole states. The obtained amorphous structures have been further fully optimised with the hybrid PBE0-TC-LRC functional.\cite{guidon2009robust} This truncated-Coulomb long-range corrected version of the hybrid functional PBE0 is known to provide accurate band gaps and structural properties of insulators and is of similar form as the hybrid HSE06 functional,\cite{krukau2006influence} but is less computationally demanding. This is achieved by truncating the computation of the exact exchange by cutoff radius ($R_c$). The amount of exact exchange and its cutoff radius can be adjusted to achieve optimum accuracy for a particular system. In this study, we use a radius cutoff $R_c \sim \SI{6}{\angstrom}$ and the amount of exact exchange $\alpha = 0.115$, which have recently been optimised \cite{Elmaslmane_TiO2_polaron18} to model polaron formation in six different \ce{TiO2} crystal structures, including rutile, anatase and brookite. Employing these parameters one obtains structural parameters of crystalline \ce{TiO2} in agreement with experiments, a band gap within $6\%$ from experimental values for anatase and rutile; and satisfies the generalised Koopmans' condition (gKc) to within \SI{0.08}{\electronvolt} for all six different \ce{TiO2} phases. The latter is important in order to provide an accurate prediction of small polarons across all \ce{TiO2} phases.\cite{Elmaslmane_TiO2_polaron18} To reduce the computational cost of nonlocal functional calculations, the auxiliary density matrix method (ADMM) was employed.\cite{guidon2010auxiliary} All geometry optimisations were performed using the BFGS optimiser to minimise forces on atoms to within \SI{0.02}{\electronvolt\per\angstrom}. All calculations are performed at the $\Gamma$ point. 

The electron/hole trapping energy, $E_T$, a measure of  stability of localised states, is calculated as the difference between total energies of the delocalised and fully localised electron states. We note that distribution of calculated $E_T$ is affected by several factors compared to similar calculations in the crystal phase, as discussed in detail in ref.~\citenum{Mora-Fonz_moiea_2020}. One of the major factors is the partially localised character of the initial state, which is discussed below along with the other aspects.
As a comparison, ten a-\ce{TiO2} structures were fully reoptimised and their trapping energies calculated  using the hybrid HSE06 functional --which has been widely used to study defects and polarons in \ce{TiO2}.

All crystal structures in this paper were generated using the VESTA package,\cite{Momma_VESTA_2011} whereas the plots have been produced using GNUPLOT.\cite{gnuplot}

\section{Results and Discussion}

\subsection{Properties of \ce{TiO2}}
First-principle calculations of \ce{TiO2} electronic structure have been discussed extensively in refs.~\citenum{chiodo2010self,kang2010quasiparticle}. 
The calculated PBE0-TC-LRC structural properties and the band gap, using $\alpha = 11.5~\%$ and a cutoff of $R_c$=\SI{6}{\angstrom}, of rutile and anatase are compared with experiment in Table~\ref{tab_cTiO2}.
For the rutile phase, the lattice parameters are $a = \SI{4.615}{\angstrom}, c = \SI{2.960}{\angstrom}$, and the band gap is \SI{2.80}{\electronvolt}, which are in good agreement with experiment (\SI{4.587}{\angstrom}, \SI{2.954}{\angstrom}, \SI{3.03}{\electronvolt}).\cite{muscat2002first,Scanlon_baora_2013}
For anatase \ce{TiO2}, the calculated lattice parameters are $a = \SI{3.788}{\angstrom}$ and $ c = \SI{9.626}{\angstrom}$ with a band gap of \SI{3.02}{\electronvolt}. These values are also in good agreement with the experimental values of $a = \SI{3.782}{\angstrom}$, $ c=\SI{9.502}{\angstrom}$ and band gap~$= \SI{3.2}{\electronvolt}$.\cite{Burdett_serii_1987,Scanlon_baora_2013}
For both \ce{TiO2} phases, the lattice parameters and band gaps are reproduced within about $1\%$ and $6\%$, respectively. We emphasise that the h-DFT functional parameters have been optimised to provide an accurate description of small polaron in a-\ce{TiO2}, which is achieved by satisfying the gKc to within \SI{0.08}{\electronvolt} for six different \ce{TiO2} crystalline phases.\cite{Elmaslmane_TiO2_polaron18}

\begin{table}[htbp]
	\centering
	\caption{Bulk properties of the rutile and anatase structure of \ce{TiO2} }
	\begin{tabular}{lccccr}
		\hline
			&Rutile & & Anatase & \\
		\hline
		& this work & expt & this work & expt \\
		&&&&\\
		$a$ (\AA)&	4.615&	4.587&	3.788&	3.782 \\
		$c$ (\AA)&	2.960&	2.954&	9.626&	9.502 \\
		Band gap (eV)&	2.80&	3.0&	3.02&	3.2 \\
		$E_T$ (h$^+$, eV)&&&			0.25&	\\
		$E_T$ (e$^-$, eV)&	0.02 &&&\\			
		\hline
	\end{tabular}%
	\label{tab_cTiO2}%
\end{table}%

\subsection{Atomic Structure of  a-\ce{TiO2}}

\begin{figure}[htpb]
	\includegraphics[width=0.99\columnwidth]{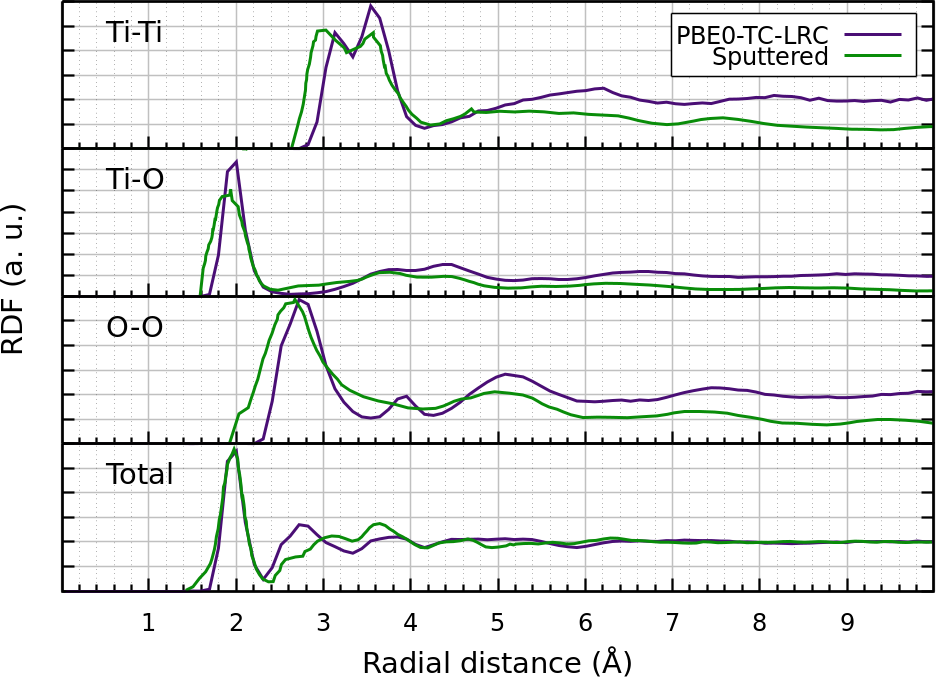}
	\caption{ Average RDF, from fifty structures, of the a-\ce{TiO2} structures fully optimised using the PBE0-TC-LRC functional. Experimental data \cite{petkov1998atomic} from sputtered \ce{TiO2} layers is shown in green as a comparison.}
	\label{fig_RDF}
\end{figure}

The topology of a-\ce{TiO2} models obtained using classical MD calculations does not change as a result of  h-DFT cell and geometry optimisation of the structures. The fully optimised PBE0-TC-LRC amorphous structures have the average density of about \SI{4.04}{\gram\per\cubic\centi\meter} ranging from 3.92 to \SI{4.14}{\gram\per\cubic\centi\meter}. 
Experimentally, a-\ce{TiO2} films have a wide range of densities and structural properties depending on preparation methods, as noted, for example, by Bendavid et al. \cite{bendavid2000deposition}, where amorphous samples obtained by filtered arc deposition range from 3.62 to \SI{4.09}{\gram\per\cubic\centi\meter} with change in the substrate bias. To our knowledge, the experimentally reported a-\ce{TiO2} densities range from 3.6 up to \SI{4.4}{\gram\per\cubic\centi\meter},\cite{fukuhara2013superior,mergel2000density,bendavid2000deposition} whereas calculations predict a value of \SI{4.18}{\gram\per\cubic\centi\meter}.\cite{deskins2017structural}
Similarly, the structures exhibit wide distributions of bond lengths, bond angles and atomic coordinations compared to the crystalline phases of \ce{TiO2}. The coordination number of each atom was determined by counting the number of atoms within a cutoff radius of \SI{2.45}{\angstrom}. The radial cutoff was chosen at the back of the first peak of the total radial distribution function (RDF) shown in Figure~\ref{fig_RDF}. The Ti ions are 5 to 7 coordinated and O ions are 2 to 4 coordinated. The average abundance (and standard deviation) 5, 6 and 7-coordinated Ti ions is about 22.78 (5.89), 67.36 (5.26) and $9.89\%$ ($3.38\%$). For the 2, 3 and 4-coordinated O atoms, the average values are 15.36 (3.19), 76.37 (3.02) and $8.28\%$ ($2.04\%$), respectively. 
We note that in r-\ce{TiO2} Ti atoms are six-fold coordinated and  oxygen atoms are three-fold coordinated. 

Our calculations also agree with the theoretical results reported previously \cite{hoang2007structural,pham2015oxygen} that use the same potential and with the experimental data.\cite{petkov1998atomic}  The RDF, averaged over the fifty fully optimised PBE0-TC-LRC structures, is shown in Figure~\ref{fig_RDF}. The main sharp pick slightly below \SI{2.0}{\angstrom} is due to the \mbox{Ti--O} bond, whereas the \mbox{O--O} pick is around \SI{2.7}{\angstrom}. The \mbox{Ti--Ti} main feature is between about 3.0 to \SI{3.8}{\angstrom} with two main picks in excellent agreement with the experimental observations,\cite{petkov1998atomic} which have been attributed to edge and corner-sharing octahedra, respectively.

\begin{figure}[htpb]
	\includegraphics[width=0.99\columnwidth]{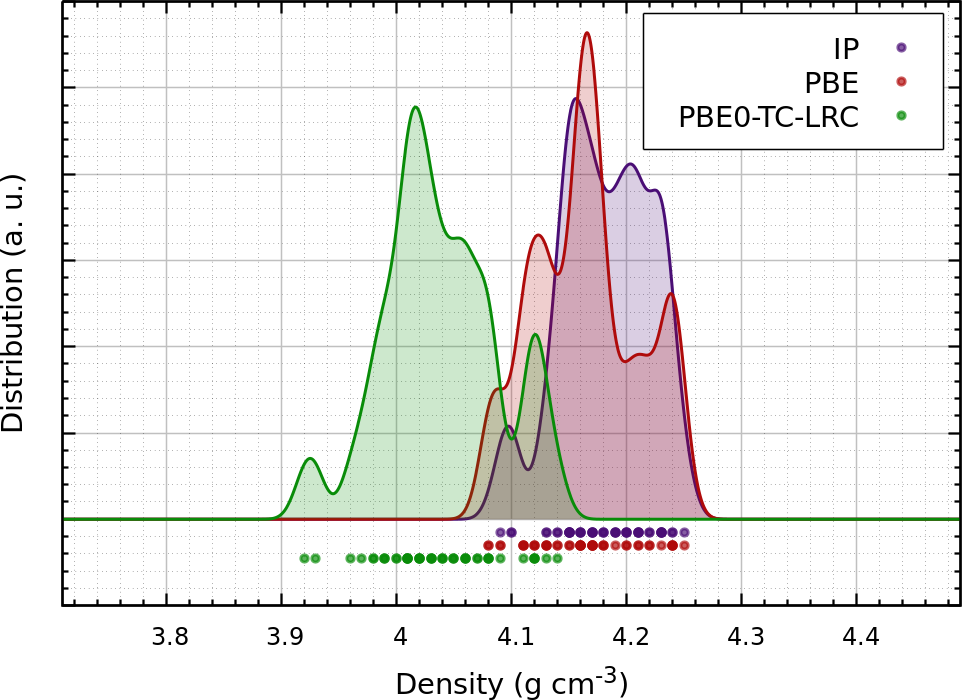}
	\caption[]{ Density distribution of fifty fully optimised a-\ce{TiO2} structures. A Gaussian smearing of $\sigma = \SI{0.01}{\gram\per\cubic\centi\meter}$ was used.}
	\label{fig_smeared_densities}
\end{figure}

\begin{figure}[htpb]
	\includegraphics[width=0.99\columnwidth]{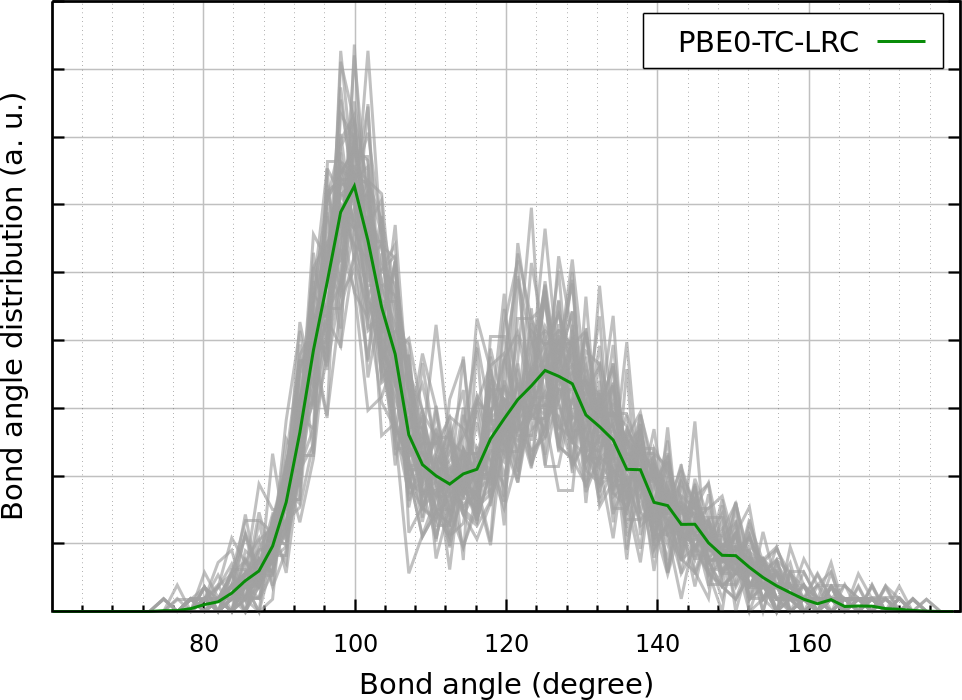}
	\caption[]{The Ti-O-Ti bond angle distribution of fifty fully optimised PBE0-TC-LRC a-\ce{TiO2} structures is shown in grey and its average in green.}
	\label{fig_angles_aTiO2}
\end{figure}

\subsection{Electronic Structure of  a-\ce{TiO2}}
\begin{figure}[htpb]
	\includegraphics[width=0.99\columnwidth]{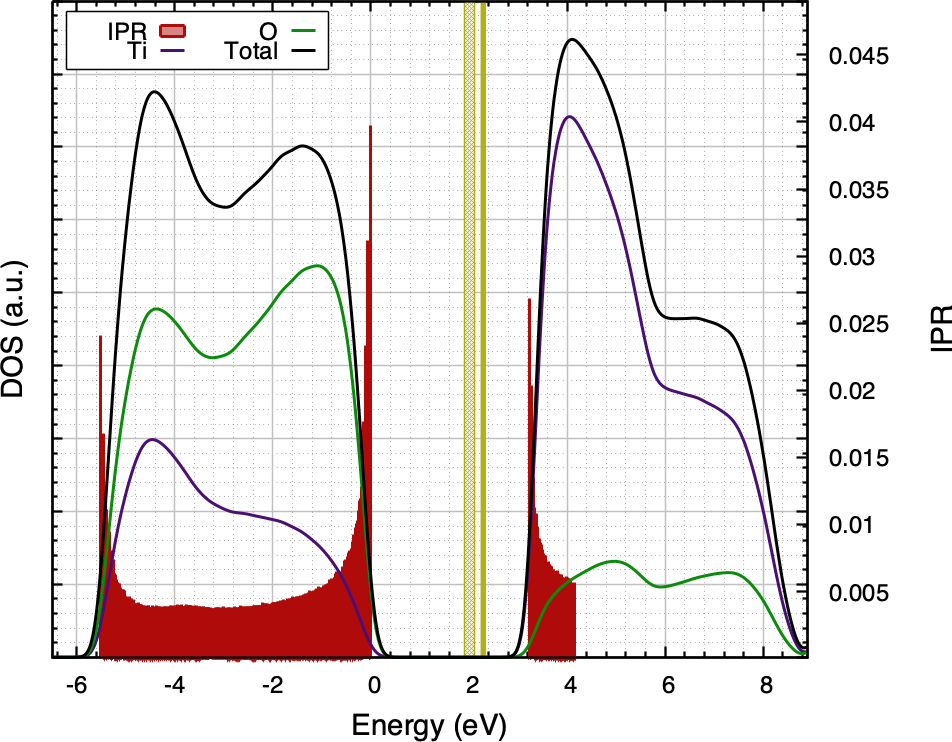}
	\caption[]{Average DoS of the a-\ce{TiO2} structures using the PBE0-TC-LRC functional and a Gaussian smearing of \SI{0.15}{\electronvolt}. The IPR values of the one-electron states quantify the charge localisation. Large IPR values at  the band edges indicate localisation of the state and small IPR values indicate delocalisation. The mobility edge position can then be estimated from the IPR spectrum. The energy ranges (average plus standard deviation) for the electron (hole) polarons in fifty amorphous structures are shown by solid (hatched) areas in the middle of the band gap. The top of the VB was set to \SI{0.0}{\electronvolt}. The a-\ce{TiO2} DoS and IPR plots were averaged from the fifty different amorphous structures.  } 
	\label{fig_DOS_IPR}
\end{figure}

Figure~\ref{fig_DOS_IPR} shows the density of states (DoS) and the Inverse Participation Ratio (IPR) spectrum of a-\ce{TiO2} calculated using h-DFT.
The a-\ce{TiO2} Kohn-Sham (KS) band gap averaged over fifty a-\ce{TiO2} structures is \SI{3.25}{\electronvolt}, with a standard deviation of \SI{0.10}{\electronvolt}. The valence band (VB) maximum consists mostly of the O 2$p$ orbitals and the conduction band edge is derived from Ti 3$d$ orbitals. The degree of localisation of these states was further analysed by calculating the IPR spectrum, which has been used to characterise the localisation of electronic states in amorphous materials including \ce{TiO2} \cite{prasai2012properties,landmann2012fingerprints} and other (more complex) amorphous structures, see 
e.g. refs.~\citenum{adelstein2015hole,Mora-Fonz_moiea_2020,Strand_ietia_2018,Dicks_tmoct_2017}. IPR is calculated for each energy eigenstate of the system and characterises its degree of localisation. The IPR formulation used here has been reported previously.\cite{Mora-Fonz_moiea_2020}  The average a-\ce{TiO2} IPR spectrum  shown in Figure~\ref{fig_DOS_IPR} is similar to those obtained in refs.~\citenum{prasai2012properties,landmann2012fingerprints}.

The IPR values for electronic states near band edges are higher, indicating that these KS states are more localised. An extra electron or hole will tend to occupy these states and structural motifs responsible for these states can be considered as precursor sites for carrier localisation. These motifs typically include 
elongated \mbox{Ti--O} bonds. 
 

\subsection{Electron and hole trapping in a-\ce{TiO2}}

\subsubsection{Single electron trapping}\label{sec_single_e_trapping}

\begin{figure*}[]
	\centering
	\begin{subfigure}[t]{0.95\columnwidth} 
		\centering
		\includegraphics[width=0.95\columnwidth]{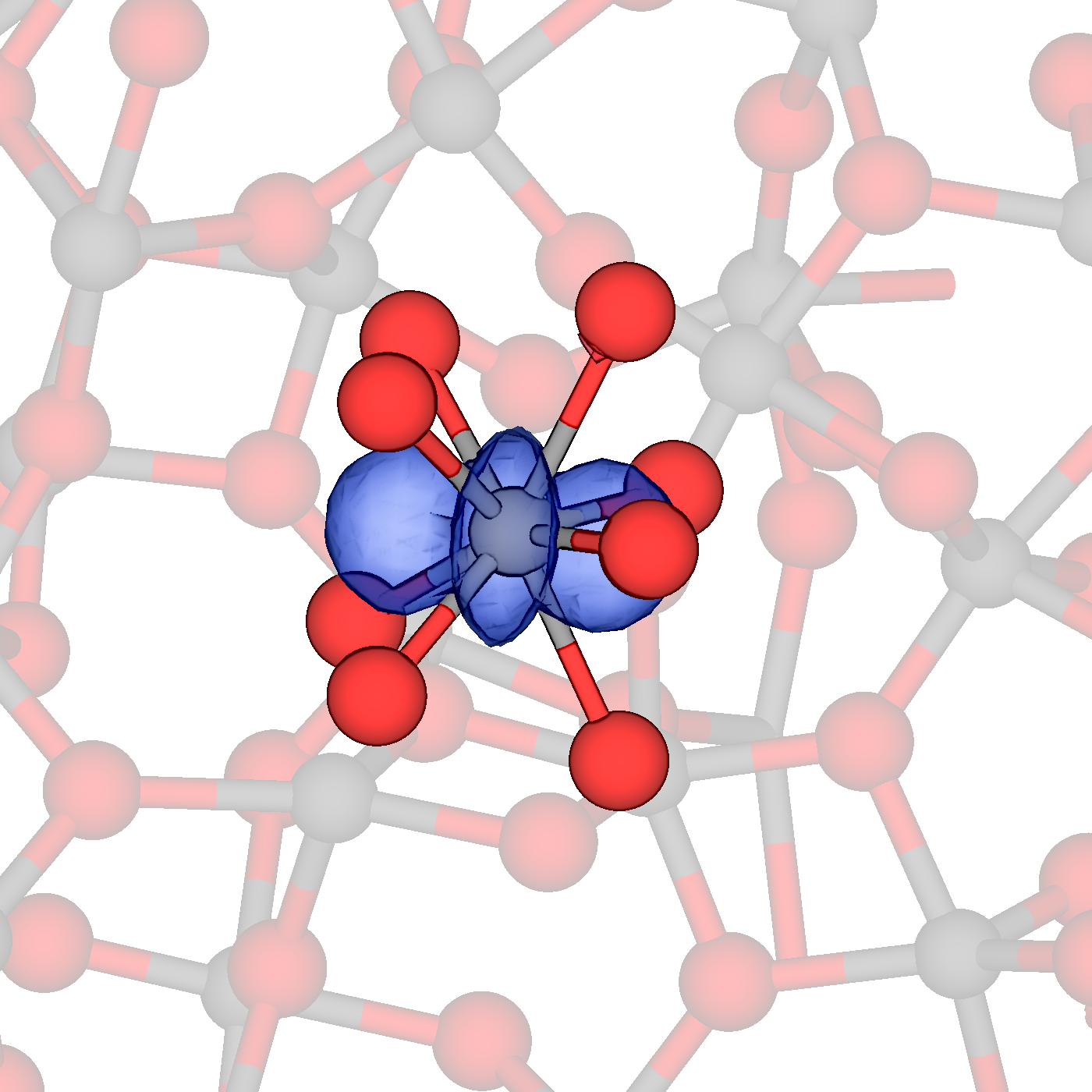}
		\caption{) \label{fig_aTiO2_electrons_1}}
	\end{subfigure}%
	~ 
	\begin{subfigure}[t]{0.95\columnwidth}
		\centering
		\includegraphics[width=0.95\columnwidth]{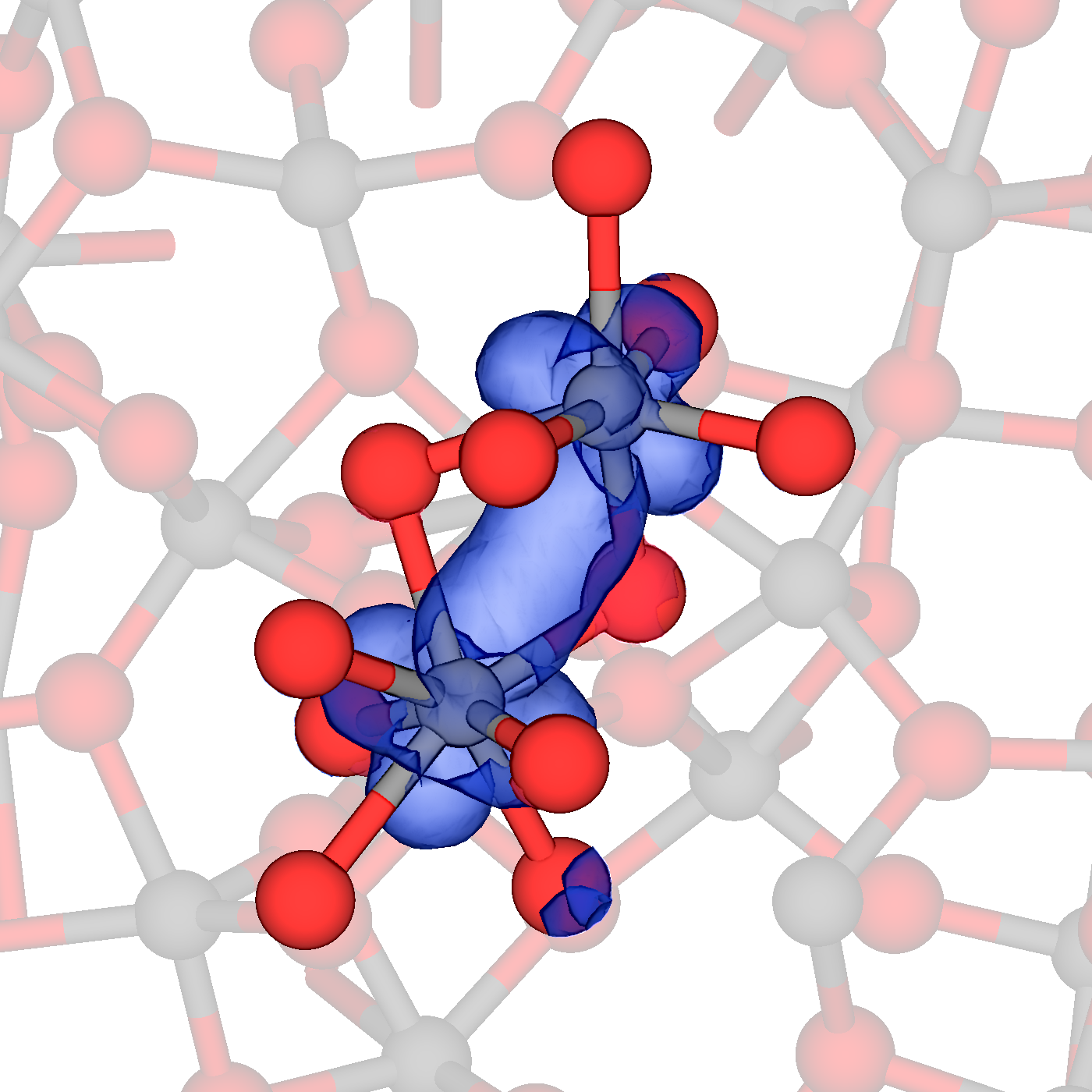}%
		\caption{) \label{fig_aTiO2_electrons_2}}
	\end{subfigure}
	\caption{ 
		The spin density of the electron polarons (blue) in a-\ce{TiO2}. a) The electron is localised on a single \ce{TiO8}  polyhedra. Ti--O and \mbox{Ti--Ti} distances are increased, on average, by about \SI{0.05}{\angstrom}. b) The electron is shared by two adjacent (\ce{TiO7} and \ce{TiO6}) polyhedra. Ti--Ti distances are reduced by \SI{0.11}{\angstrom}, whereas Ti--O distance is increased by \SI{0.03}{\angstrom}, on average, for both Ti ions holding the electrons. The electron always occupies Ti $3d$ orbitals. Gray and red colours are reserved for Ti and O ions, respectively.
	}
	\label{fig_aTiO2_electrons}
\end{figure*}

To investigate the electron trapping further, an extra electron ($N+1$, where $N$ is the number of electrons for the neutrally charge structure) was added in fifty amorphous structures. As can be expected from the IPR analysis, initial electron states are not completely delocalised over the entire cell, but rather exhibit localisation on few Ti ions. The degree of initial electron localisation across all the samples is quite wide. Upon the geometry optimisation, every structure showed further spontaneous electron localisation on Ti $3d$ orbitals. We note that most of the electrons localise on 6-coordinated Ti ions, whereas there are few cases involving 7-coordinated Ti ions and a very small number on 8-coordinated Ti ions (as the one shown in Figure~\ref{fig_aTiO2_electrons_1}). 

Broadly, we can identify two different electron localisation types, shown in Figure~\ref{fig_aTiO2_electrons}, which are present in similar concentrations. In the first type (Figure~\ref{fig_aTiO2_electrons_1}), most of the spin density is localised on a single Ti ion, with  \mbox{Ti--O} and \mbox{Ti--Ti} distances increased by about \SI{0.05}{\angstrom} on average.
The latter refers to local distortions close to the localised site (e.g. first nearest neighbours).
For the second type (Figure~\ref{fig_aTiO2_electrons_2}), the spin density is localised within two corner-sharing Ti polyhedra. In most cases the extra electron is shared between the two Ti ions (as shown in Figure~\ref{fig_aTiO2_electrons_2}) with the distance between them reduced by about \SI{0.11}{\angstrom}, whereas the \mbox{Ti--O} bonds are elongated by \SI{0.03}{\angstrom}, on average. In few remaining configurations the extra electron sits on two or three  adjacent Ti ions with no significant spin density overlap between them. 

Trapped electrons create deep KS states in the band gap located at $\sim$\SI{0.96}{\electronvolt} below the bottom of the CB with a standard deviation of \SI{0.11}{\electronvolt} as indicated by the solid area in Figure~\ref{fig_DOS_IPR} with the width of the area corresponding to the standard deviation of distribution of occupied electron states. This behaviour differs from the case of rutile and anatase where electron polarons form shallow states on regular Ti sites.

\begin{figure}
	\centering
	\includegraphics[width=0.98\columnwidth]{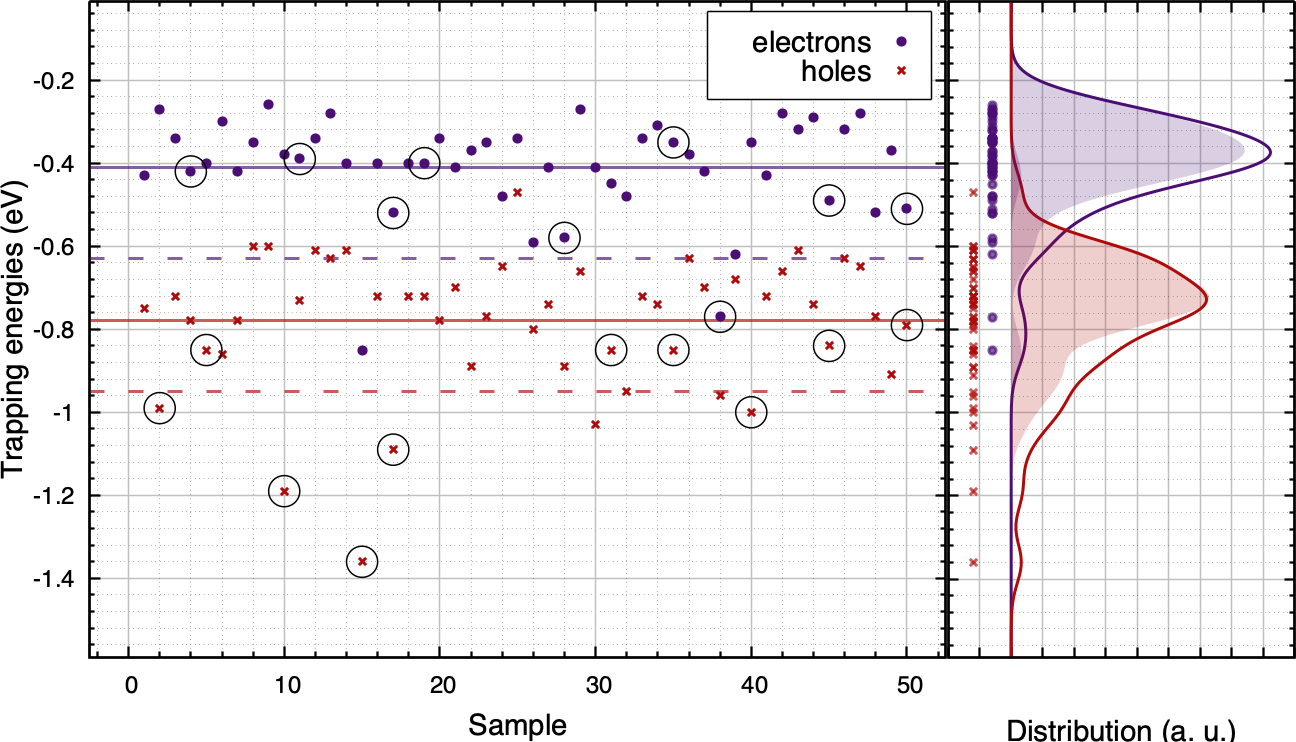}
	\caption{Calculated $E_T$ for fifty 270-atom a-\ce{TiO2} structures using the PBE0-TC-LRC  functional with $\alpha = 11.5~\%$ and a cutoff of $R_c$=\SI{6}{\angstrom}.  Horizontal lines represent the averages. The dashed horizontal lines show the average, taken from ten 270-atom a-\ce{TiO2} structures, using the HSE06 functional. The irreversible cases are shown encircled and the filled curves represent the $E_T$ distribution of without these metastable structures.  The smeared $E_T$ distribution is shown on the right with a $\sigma = \SI{0.05}{\electronvolt}$.}
		\label{fig_trapping_energies}
	\end{figure}

The electron trapping energies in fifty a-\ce{TiO2} models were calculated as the total energy differences between the initial (partially localised) electron state in the amorphous structure and after the geometry optimisation (see Figure~\ref{fig_trapping_energies}). The average $E_T$ is about \SI{-0.4}{\electronvolt} with a wide distribution ranging between \SI{-0.26}{\electronvolt} and \SI{-0.85}{\electronvolt} and a standard deviation of \SI{-0.12}{\electronvolt}. The latter corresponds to the standard deviation width of the defect KS state levels created in the band gap by the localised electron, shown in Figure~\ref{fig_DOS_IPR}. 
 We note that these $E_T$ are much deeper than those calculated in r-\ce{TiO2} (\SI{-0.02}{\electronvolt}) using the same method, no electron localisation was predicted for the anatase structure.
However, for more accurate determination of $E_T$ and comparison with those in crystal structures one should use the $E_T$ values calculated with respect to the delocalised states located above the mobility edge. Using the IPR analysis in Figure~\ref{fig_DOS_IPR} one can estimate that the electron mobility edge is located about \SI{0.4}{\electronvolt} deeper in the CB, therefore, $E_T$ should be closer to \SI{-0.80}{\electronvolt} for electrons. The large value of $E_T$ suggests stability of localised electrons at room temperature. 
Moreover, as we show in Figure~\ref{fig_DOS_IPR}, the use of the hybrid HSE06 functional  does not change our qualitative conclusions.

The IPR analysis (Figure~\ref{fig_DOS_IPR}) demonstrates that, on average, there are 2-3 precursor sites per 270 atoms for an electron to trap. These can be explored either by adding more electrons to the cell or by inducing structural distortion near precursor sites to facilitate the electron localisation. In contrast with a-ZnO,\cite{Mora-Fonz_moiea_2020} we could not find a clear correlation between the $E_T$ and the number of Ti ions holding the spin density.

\subsubsection{Single hole trapping}\label{sec_single_h_trapping}
Hole localisation was studied by removing one electron from the system ($N-1$). In r-\ce{TiO2} we  find no  hole trapping but in anatase our calculations predict the hole $E_T$ of $\sim$\SI{-0.25}{\electronvolt}. The average IPR spectrum in Figure~\ref{fig_DOS_IPR} suggests hole localisation in a-\ce{TiO2} with potentially more precursor sites (3-4) than for the electron trapping (2-3). 

\begin{figure*}[]
	\centering
	\begin{subfigure}[t]{0.95\columnwidth} 
		\centering
		\includegraphics[width=0.95\columnwidth]{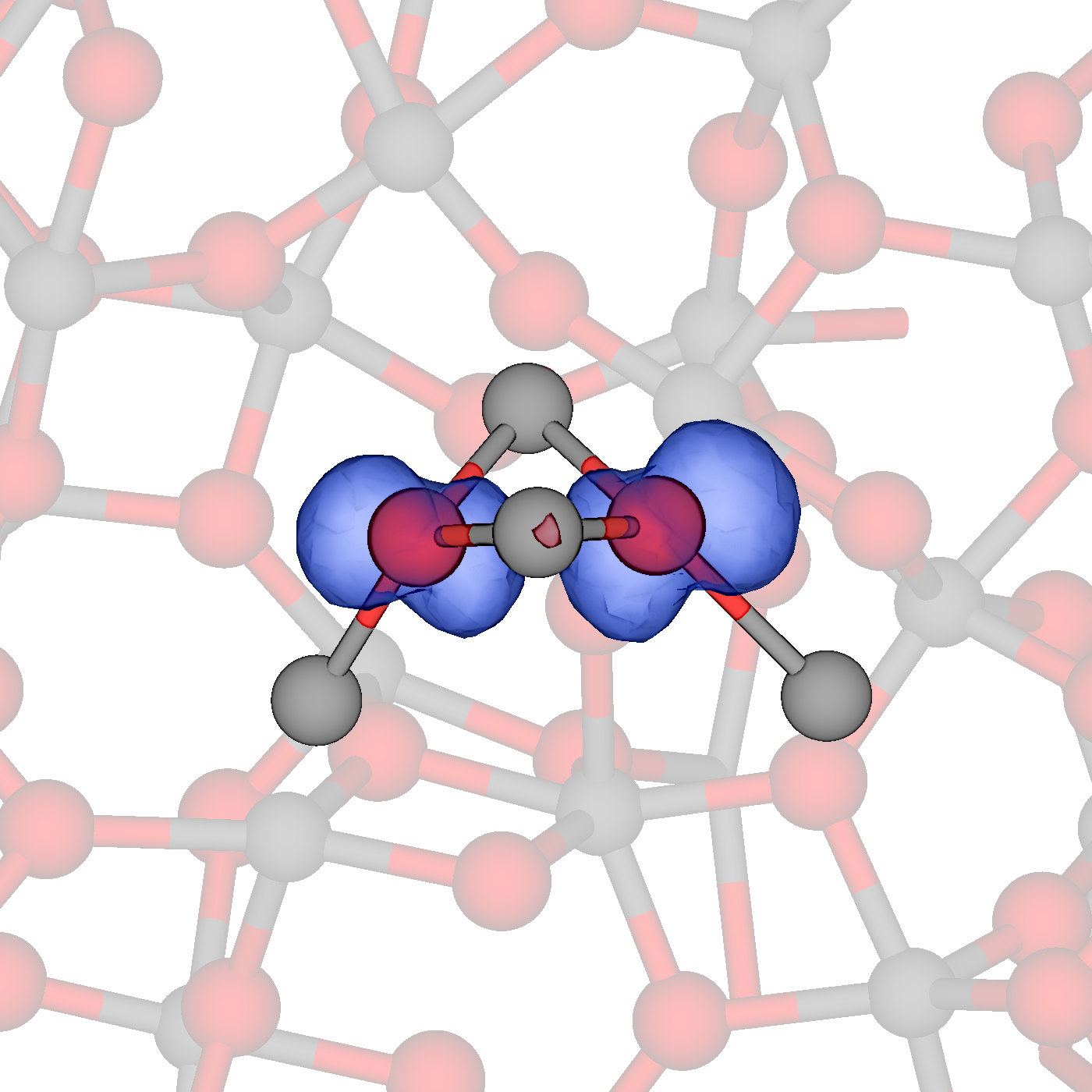}
		\caption{) \label{fig_aTiO2_holes_1}}
	\end{subfigure}%
	~ 
	\begin{subfigure}[t]{0.95\columnwidth}
		\centering
		\includegraphics[width=0.95\columnwidth]{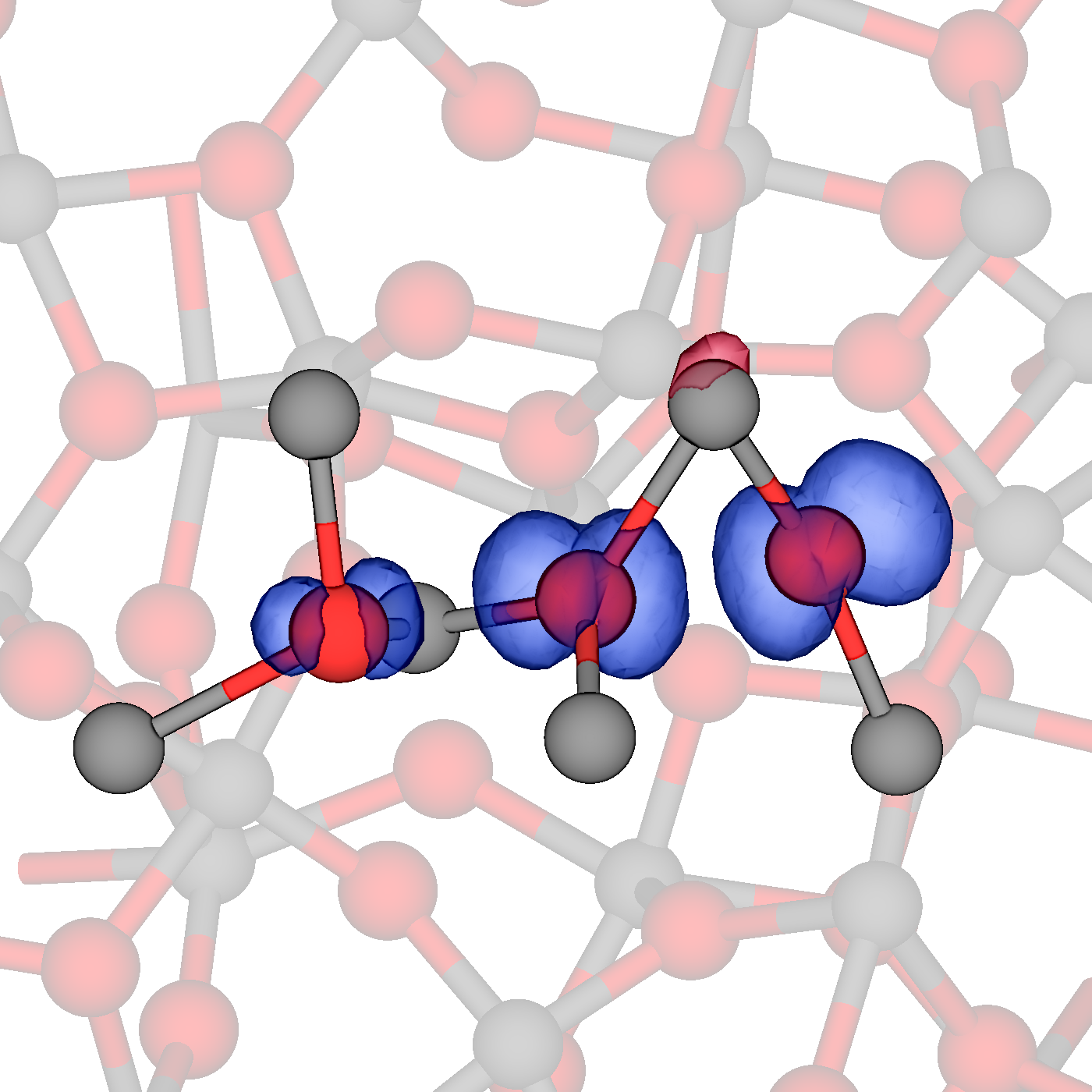}%
		\caption{) \label{fig_aTiO2_holes_2}}
	\end{subfigure}
	\caption{ 
		The spin density of the hole polarons (blue) in a-\ce{TiO2}. a) The hole is shared by two adjacent 3-coordinated O ions. O--O distances are $12\%$ shorter than for the neutral case, whereas Ti--O distances are larger by $3.6\%$, on average.  b) The hole is localised in three adjacent O atoms. On average, Ti--O distances increase by $2.5\%$ and decrease by about $8.8\%$ for O--O distances.    The hole always occupies O $2p$ orbitals. Note that the geometry of these features is planar-like in every case. 
	}
	\label{fig_aTiO2_holes}
\end{figure*}

Every a-\ce{TiO2} structure exhibits hole trapping states on O $2p$ orbitals. There are two types of localised hole states: in the first one the hole is localised on two adjacent O ions (Figure~\ref{fig_aTiO2_holes_1}), whereas in the second the hole is shared by three O atoms (Figure~\ref{fig_aTiO2_holes_2}).
Similar to a-ZnO,\cite{Mora-Fonz_moiea_2020} the geometry of these hole states is planar-like in every structure, as seen in Figure~\ref{fig_aTiO2_holes}. Among the fifty structures, we did not find one where the hole is localised solely on one O ion. Hole localisation causes stronger network distortion than that induced by electrons. One of the reasons for this is that, on average, the states at the top of the VB are more localised than those at the bottom of the CB (see Figure~\ref{fig_DOS_IPR}). The \mbox{Ti--O} (O--O) distances are elongated (reduced) after the hole localisation by ca. \SI{0.09}{\angstrom} (\SI{0.25}{\angstrom}) on average. 



The unoccupied holes states after the geometry optimisation are located approximately \SI{2.0}{\electronvolt} above the top of the VB (\SI{1.25}{\electronvolt} below the CB) and a standard deviation of \SI{0.19}{\electronvolt} (see hatched area in Figure~\ref{fig_DOS_IPR}). We note that Pham et al.~\cite{pham2015oxygen} predicted the hole polaron KS state  at $\sim$\SI{0.6}{\electronvolt} above the VBM, which is outside of the distribution obtained in this work ranging from \SI{1.51}{\electronvolt} to \SI{2.42}{\electronvolt} in the fifty amorphous structures (see Figure~\ref{fig_DOS_IPR}). We believe that this discrepancy is mainly due to the underestimation of the band gap and the different description of charge localisation resulting from the DFT+$U$ formalism used in ref.~\citenum{pham2015oxygen} compared to the hybrid exchange correlation functional. 

The average hole $E_T$ is \SI{ -0.78}{\electronvolt}, ranging between \SI{-0.37}{\electronvolt} and \SI{-1.36}{\electronvolt}. These values are again much deeper than those found for hole trapping in anatase \ce{TiO2} (\SI{-0.25}{\electronvolt}). When compared to electron traps in a-\ce{TiO2}, $E_T$ in hole polarons are about twice deeper.
For holes, the estimated mobility edge is about \SI{0.5}{\electronvolt} below the VBM, which would increase the $E_T$ for holes to \SI{1.28}{\electronvolt}, on average. 
Similar to localised electrons, there is no change in our qualitative results (see Figure~\ref{fig_DOS_IPR}) using hybrid HSE06 functional.

\subsubsection{Reversibility of localised states}

As described above, the electron and hole localisation causes significant distortions of the surrounding amorphous \ce{TiO2} network. This may cause irreversible changes in model amorphous structure and its relaxation can be part of trapping energies reported in Figure~\ref{fig_trapping_energies}. When the missing charge is injected back into an amorphous structure, it does not always return into its initial state but may transfer into another (lower) minimum of the potential energy landscape. Computationally, this process corresponds to e.g. photo-induced ionisation of trapped states accompanied by  relaxation of the neutral structure. This is akin to well-known photo-induced structural changes in amorphous solids particularly well studied in chalcogenides and amorphous H:Si (see e.g. \cite{Mikla1996photo,Tanaka1998photo}).

For trapped electrons (holes), we found $9/50$ ($10/50$) structures with a total energy difference between the two neutral states greater than \SI{0.1}{\electronvolt} and a total energy gain, on average, of \SI{0.26}{\electronvolt} (\SI{0.23}{\electronvolt}). This total energy difference may seem small for a 270-atom unit cell. To find out whether the resulting structure is a new local minimum rather than a result of numerical errors, we searched for ionic displacements greater than \SI{0.2}{\angstrom} in each case. We conclude that in 18 out 19 cases a new lower minimum was found. On average, for electron traps, there are eight ions moving by \SI{0.32}{\angstrom}, whereas for hole traps, there are 10 ions that move by \SI{0.30}{\angstrom}. In some cases, even stronger ionic displacements than \SI{0.75}{\angstrom} are induced. The topology of the two local minima is, however, the same. In Figure~\ref{fig_trapping_energies} we highlight the  trapping energies of ``irreversible" structures and the $E_T$ distribution without these metastable ``irreversible" cases (shown with filled curves), which is normal-like without noticeable tails. One can see that the $E_T$ outliers for hole traps correspond to strong network relaxation leading to transition into new structures. 

\subsection{Electron and hole bipolaron-like states in a-\ce{TiO2}}

The interaction between localised electrons or holes can lead to formation of more stable systems, often called bipolaron. Hole bipolarons (double holes) have been suggested in many crystalline metal oxides including anatase \ce{TiO2}, \ce{MoO3}, \ce{V2O5}, \ce{InGaZnO}, \ce{HfO2}, and others.\cite{Meux_eohst_2018,Nahm_ioaos_2012,Chen2014doublehole,Strand2017holes} Similarly, the existence of electron bipolarons was predicted in a-\ce{SiO2},\cite{gao2016mechanism}, where Si--O--Si precursor sites act as deep electron traps and can accommodate up to two extra electrons, and in a-\ce{HfO2}.\cite{kaviani2016deep} In this study, we have investigated the interaction of localised electrons and holes in a-\ce{TiO2} by adding an extra electron (hole) to the existing localised electron (hole) structures in fifty samples. We will call the resulting states bipolarons for brevity in full realisation that their stability is caused to a very significant extent by the structural disorder.

In most cases polarons prefer to stay apart, unless the constituent atoms form bonds or there is a favourable interference of lattice distortions caused by polarons. A measure of the interaction between bipolarons in a crystal is given by its binding energy, $E_{bind}$, which is defined as:
\begin{equation} \label{eq_binding}
    E_{bind} = 2E_{polaron} - [ E_{bipolaron} + E_{neutral} ],
\end{equation}
where $E_{polaron}$ ($E_{bipolaron}$) and  $E_{neutral}$ are the charge corrected energies of the geometry optimised single (double) polaron and neutral periodic cell, respectively. Positive values indicate, therefore, a higher stability for the bipolaron with respect to the two identical infinitely separated polarons. In the amorphous phase, all sites are different and, hence, this expression is approximate. Here we use the energy of the most stable single localised state in each amorphous sample and that obtained after adding the second electron or hole.

\subsubsection{Interaction between localised electrons in a-\ce{TiO2}}
As suggested by the IPR analysis and DoS (Figure~\ref{fig_DOS_IPR}), extra electrons in a-\ce{TiO2} localise in Ti $3d$ molecular orbitals. Bipolarons are formed either by bonding two Ti ions or in Ti polyhedra sites that can accommodate up to two extra electrons, similar to the structures shown in Figure~\ref{fig_aTiO2_electrons}.
Most of the density of the extra electrons is localised on about 2-3 cations.  The average distance between the two cations is about \SI{3}{\angstrom} (with the shortest being ca. \SI{2.75}{\angstrom}), which corresponds to the shortest distance of the first Ti--Ti nearest neighbours seen in  neutral structures (see Figure~\ref{fig_RDF}).  When compared to the $N+1$ case, there is significant further relaxation on the next nearest neighbours when the second electron is added.

\begin{figure*}[]
	\centering
	\begin{subfigure}[t]{0.95\columnwidth} 
		\centering
		\includegraphics[width=0.95\columnwidth]{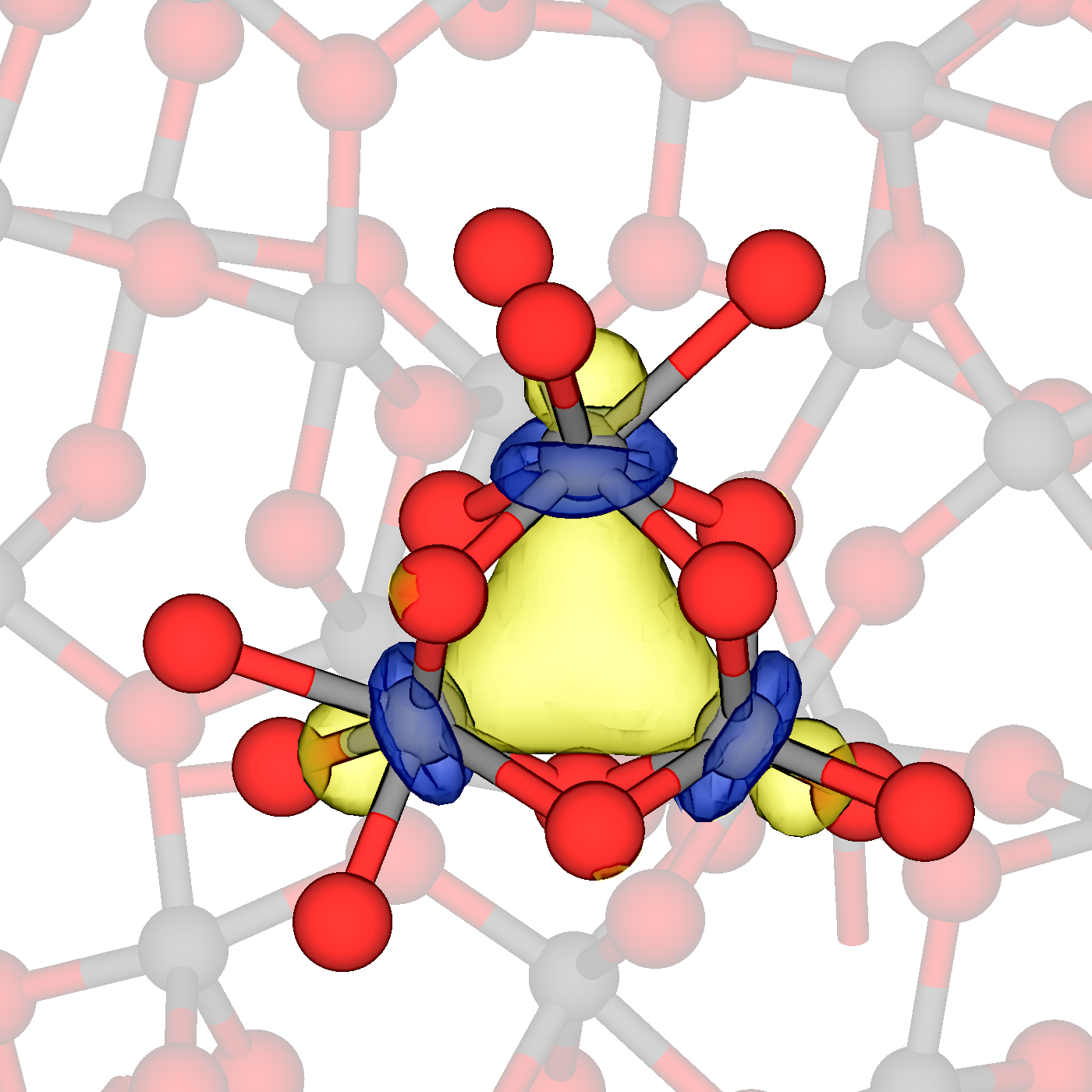}
		\caption{) \label{fig_aTiO2_bi-electrons_3}}
	\end{subfigure}%
	~ 
	\begin{subfigure}[t]{0.95\columnwidth}
		\centering
		\includegraphics[width=0.95\columnwidth]{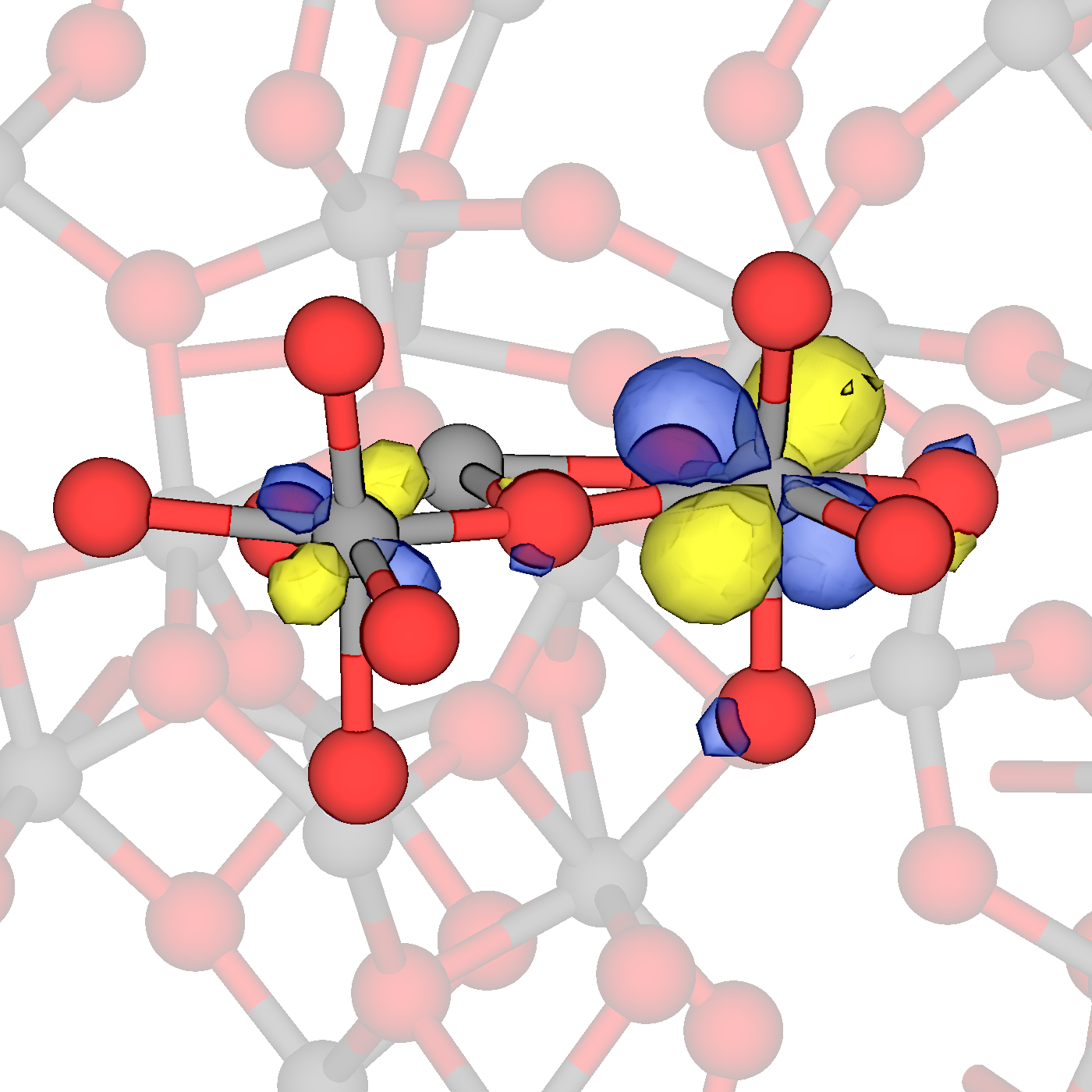}%
		\caption{) \label{fig_aTiO2_bi-electrons_4}}
	\end{subfigure}
	\caption{ 
		Localisation of two electrons in a-\ce{TiO2}. a) Type (iii), electrons are shared by three adjacent Ti polyhedra. b) Type (iv), electrons are localised by two Ti ions in a planar-like Ti-O-Ti structure.  The coloured surfaces correspond to the unoccupied KS states, with blue (yellow) indicating a positive (negative) isovalue. The magnitude of the isovalue was set to 0.05.
	}
	\label{fig_aTiO2_bi-electrons}
\end{figure*}

We can distinguish four structural types of doubly negatively charged structures: (i) the most common is two corner-sharing Ti polyhedra, with two electrons shared mostly between two Ti ions (in ca. $30\%$ of the cases); this interaction leads to the shorter Ti--Ti distances (similar to Figure~\ref{fig_aTiO2_electrons_2}); (ii) in about $16\%$ of the structures, the bipolaron is localised mostly on one Ti ion (similar to Figure~\ref{fig_aTiO2_electrons_1}); (iii) bipolaron is delocalised over three Ti ions (Figures~\ref{fig_aTiO2_bi-electrons_3}); and (iv) in fewer cases, planar Ti-O-Ti connections are formed as illustrated in Figure \ref{fig_aTiO2_bi-electrons_4}. We observe that in a third of the cases where two electrons are localised on one Ti ion this is not the same Ti where the first electron was sitting but a different precursor site. In other words, in the process of convergence a different stable site is found. Whether this observation has any significance in terms of bipolaron mobility requires separate investigation. 

\begin{figure}
	\centering
	\includegraphics[width=0.98\columnwidth]{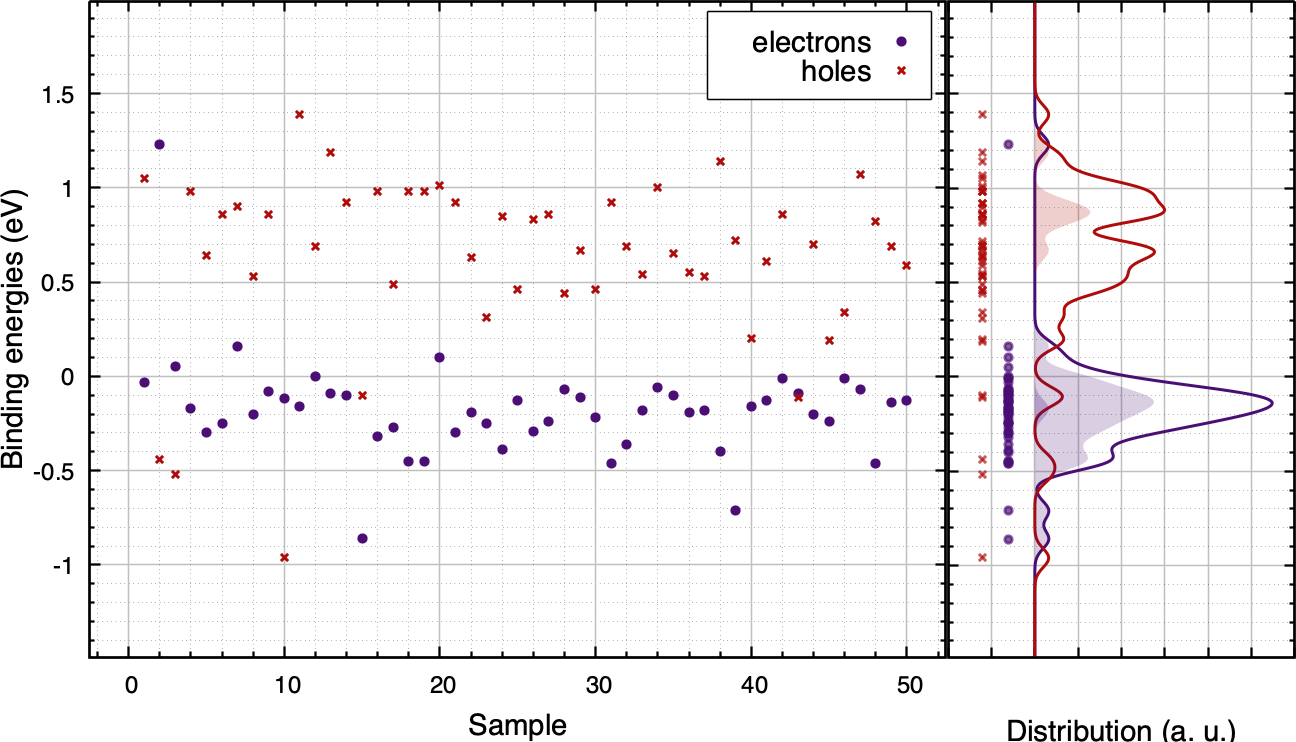}
	\caption{Calculated binding energies for fifty 270-atom a-\ce{TiO2} structures using the PBE0-TC-LRC  functional with $\alpha = 11.5~\%$ and a cutoff of $R_c$=\SI{6}{\angstrom}. The filled curves represent the distribution of bipolaron states. The smeared binding energy distribution is shown on the right with a $\sigma = \SI{0.05}{\electronvolt}$.}
		\label{fig_binding_energies}
\end{figure}

The distribution of bipolaron binding energies is shown in Figure~\ref{fig_binding_energies}. In general, two separated polarons are more stable than a bipolaron.  Only few structures show a higher stability than two singly charged configurations. The latter is caused by further geometry relaxation due to the electron injection. The energies of bipolaron  states (types (i) and (ii), see filled curves in Figure~\ref{fig_binding_energies}) are distributed across the whole range.  The distribution of the binding energies is homogeneous, with energies going from \SI{0.0}{\electronvolt}  to \SI{-0.5}{\electronvolt} in most cases. 

\subsubsection{Interaction between localized holes in a-\ce{TiO2}}
\begin{figure*}[]
	\centering
	\begin{subfigure}[t]{0.95\columnwidth} 
		\centering
		\includegraphics[width=0.95\columnwidth]{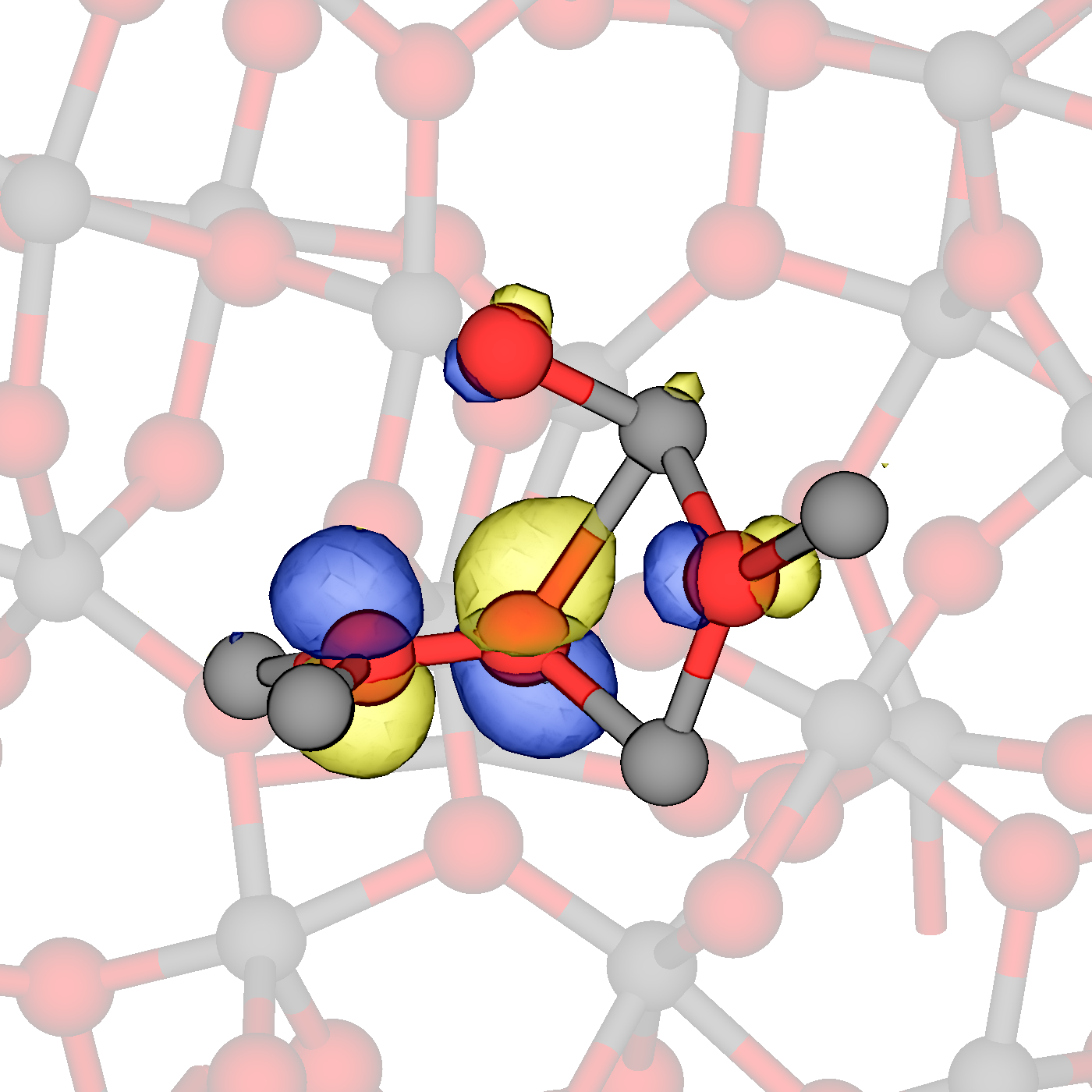}
		\caption{) \label{fig_aTiO2_bi-holes_1}}
	\end{subfigure}%
	~ 
	\begin{subfigure}[t]{0.95\columnwidth}
		\centering
		\includegraphics[width=0.95\columnwidth]{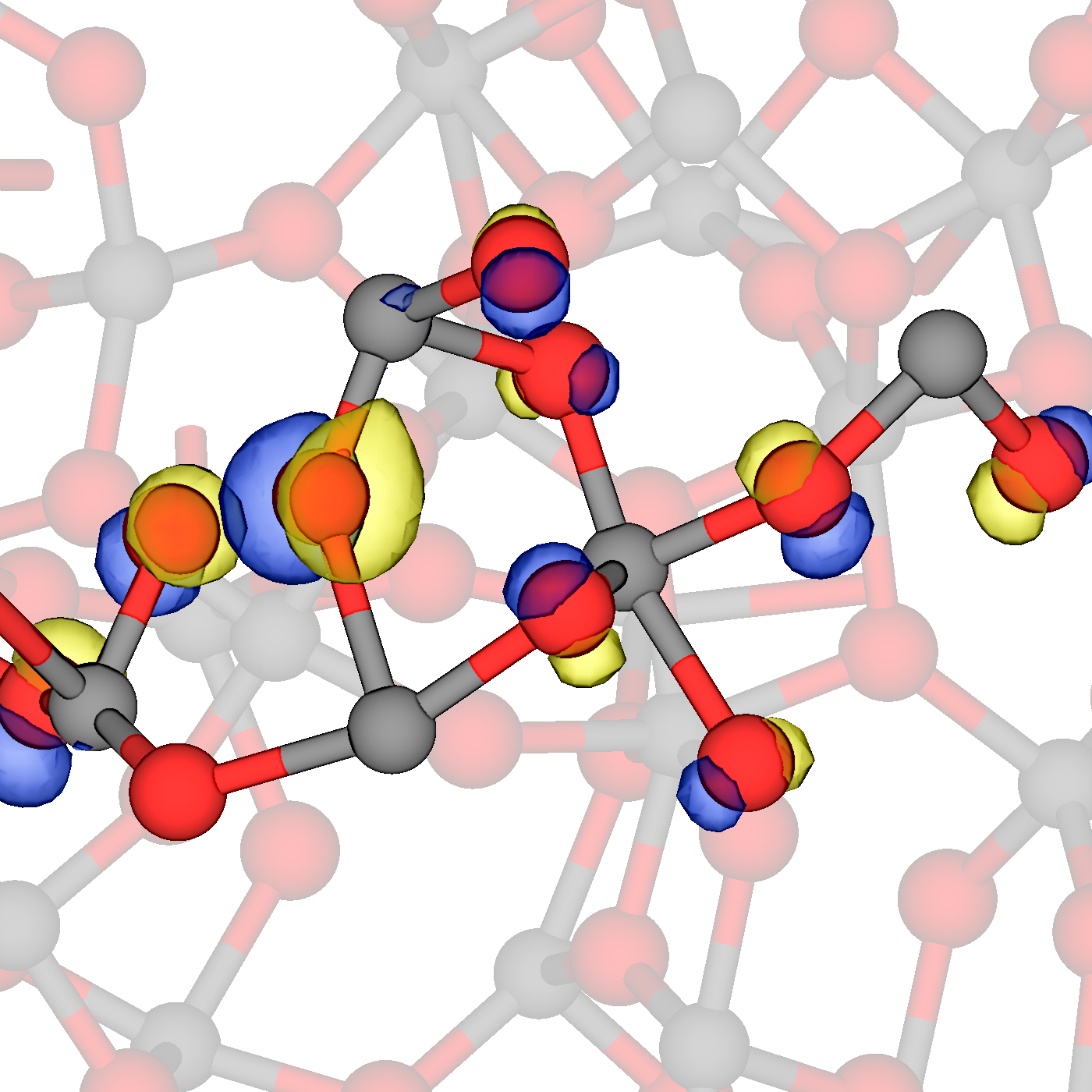}%
		\caption{) \label{fig_aTiO2_bi-holes_2}}
	\end{subfigure}
	\caption{ 
		Localisation of two holes in a-\ce{TiO2}. a) Hole bipolaron. b) Unoccupied KS states are localised on adjacent O ions forming a linear-like localisation that runs across the lattice. The magnitude of the isovalue was set to 0.05.
	}
	\label{fig_aTiO2_bi-holes}
\end{figure*}

As mentioned above, the formation of hole-bipolaron states, which are associated with the formation of peroxide-like O--O bonds, has been reported for anatase \ce{TiO2} and other oxides. For a-\ce{TiO2}, we have found that in $12\%$ (1 bipolaron for every 2250 atoms) of the structures, there is no energetic barrier to create hole bipolarons (see Figure~\ref{fig_aTiO2_bi-holes_1}). The O--O distance between ions holding the holes is about 1.43~-~\SI{1.45}{\angstrom}, which agrees with those reported in anatase \ce{TiO2} (1.45~-~\SI{1.49}{\angstrom}).\cite{Chen2014doublehole}
We note that the wave function of the unoccupied hole state does not only sit on the the O--O pair but also on neighbouring O ions. This reduces the Coulomb repulsion between the positive charges facilitating the O--O formation. Hole bipolarons can be formed with O ions of the same Ti polyhedra or through Ti--O--O--Ti connections.

For the remaining configurations, the two holes are localised on adjacent O ions about, on average, \SI{2.47}{\angstrom} away from each other forming either linear-like localisation as shown on Figure~\ref{fig_aTiO2_bi-holes_2} or clustering the localisation. Similar for the single hole polarons, the unoccupied KS states are shared by several O atoms. In about $80\%$ of the $N-2$ structures, the polaron sites differ from those on the $N-1$ cases, which suggests that creation of new precursor sites can be achieved through small geometry distortions. We also observe that hole bipolarons are significantly more delocalised than two electrons, which is particularly evident in Figure~\ref{fig_aTiO2_bi-holes_2}.

In contrast to the electron doubly charged structures, hole bipolarons are, in general, more stable than two single polarons, with a binding energies between \SI{0.2}{\electronvolt}  to \SI{1.15}{\electronvolt} in most cases. There is some correlation between the binding energies and the bipolaron formation, with bipolarons being, on average, more stable than cases with two separated polarons. 
The energies of bipolaron  states (Figure~\ref{fig_aTiO2_bi-holes_1}, see filled curves in Figure~\ref{fig_binding_energies}) are distributed across the whole range.

\section{Summary and Conclusions}
To summarise, we studied intrinsic electron and hole trapping in pure amorphous \ce{TiO2} structures. Our results demonstrate that for a-\ce{TiO2}, both electrons and holes can  be trapped at precursor sites in deep gap states. We have identified these precursor sites by using the IPR spectrum and analysis of trapped states as elongated \mbox{Ti--O} bonds. The electron localisation leads to the formation of localised states with energies about \SI{0.96}{\electronvolt} below the bottom of the CB and $E_T$ about \SI{-0.4}{\electronvolt}. The hole $E_T$ are even deeper at around \SI{-0.8}{\electronvolt} with localised states at around \SI{2}{\electronvolt} above the top of the VB. With the caveats of the density functional used the results demonstrate that, similar to other oxides, the electron and hole localisation in amorphous \ce{TiO2} creates much deeper states than in crystalline phases. The similar strong electron localisation takes place at surfaces and in nanocrystals, where the electrostatic potential and ion coordination near the surface play a crucial role in trapping the extra electrons and holes.\cite{deskins2009localized,wallace2015facet,tabriz2017application,selccuk2017excess} However, the local disorder of the amorphous structures amplifies the polaronic relaxation and $E_T$. Our results demonstrate that a-\ce{TiO2} combines the charge trapping properties of both rutile and anatase with the electron (hole) $E_T$ at precursor sites being much larger in the amorphous structures. The results can be used for understanding the mechanisms of photo-catalysis and improving the performance of electronic and memory devices employing a-\ce{TiO2} films.

One of the main effects of the deep electron and hole trapping at precursor sites in a-\ce{TiO2} is on the carrier mobility. The disordered nature of amorphous materials usually leads to percolative carrier transport with a large characteristic length scale. Accurate simulation of this transport is still a challenging problem for DFT calculations (see e.g. ref.~\citenum{masse2016ab}) which goes beyond the scope of this work. Electron transport in a-\ce{TiO2} should involve tunnelling between deep precursor sites and thermal activation into ``regular" network sites and hopping between those sites. This may represent an interesting case of crossover from Mott to Efros-Shklovskii variable-range-hopping conductivity discussed in e.g. refs.~\citenum{rosenbaum1991crossover,castner1991hopping} for other oxide films.

\begin{acknowledgments}
D.M.-F. and A.L.S. acknowledge funding provided by EPSRC under grants EP/K01739X/1 and EP/P013503/1, and by the Leverhulme Trust grant RPG-2016-135. 
M.K. and A.L.S. are grateful to the World Premier International Research Center Initiative (WPI) sponsored by the Ministry of Education, Culture, Sports, Science and Technology (MEXT), Japan for financial support.
Computer facilities on the ARCHER UK National Supercomputing Service have been provided via the UKs HPC Materials Chemistry Consortium (EPSRC Grant No. EP/L000202). This work used the  UK Materials and Molecular Modelling Hub for computational resources, which is partially funded by EPSRC (EP/P020194). The authors acknowledge the use of the UCL Grace High Performance Computing Facility (Grace@UCL), and the associated support services, in the completion of this work. The authors wish to thank Mohammad Koleini for help in calculations and valuable discussions.
\end{acknowledgments}


\bibliography{main_PRB.bib}

\end{document}